\documentclass[a4paper,conference]{IEEEtran}
\IEEEoverridecommandlockouts
\usepackage{cite}
\usepackage{amsmath,amssymb,amsfonts}
\usepackage{algorithmic}
\usepackage{graphicx}
\usepackage{textcomp}
\usepackage{xcolor}
\usepackage{xurl}
\usepackage[OT1]{fontenc} 
\usepackage{hyperref}
\def\BibTeX{{\rm B\kern-.05em{\sc i\kern-.025em b}\kern-.08em
    T\kern-.1667em\lower.7ex\hbox{E}\kern-.125emX}}
\begin{document}
	
	\begin{titlepage}
	\begin{center}
		
		\Huge
		\textbf{Anchor Pair Selection in TDOA Positioning Systems by Door Transition Error Minimization		
		}
		
		\vspace{0.5cm}
		\LARGE
		Accepted version
		
		\vspace{1.5cm}
		
		\text{Marcin Kolakowski, Jozef Modelski}
		
		\vspace{.5cm}
		\Large
		Institute of Radioelectronics and Multimedia Technology
		
		Warsaw University of Technology
		
		Warsaw, Poland,
		
		contact: marcin.kolakowski@pw.edu.pl

		\vspace{2cm}

	\end{center}
	
	\Large
	\noindent
	\textbf{Originally presented at:}
	
	\noindent
2022 24th International Microwave and Radar Conference (MIKON), Gdansk, Poland, 2022
	
	\vspace{.5cm}
	\noindent
	\textbf{Please cite this manuscript as:}
	
	\noindent
M. Kolakowski and J. Modelski, "Anchor Pair Selection in TDOA Positioning Systems by Door Transition Error Minimization," 2022 24th International Microwave and Radar Conference (MIKON), Gdansk, Poland, 2022, pp. 1-4, doi: 10.23919/MIKON54314.2022.9924777.
	
	\vspace{.5cm}
	\noindent
	\textbf{Full version available at:}
	
	\noindent
	\url{https://doi.org/10.23919/MIKON54314.2022.9924777}

	%
	%
	
	\vfill
	
	\large
	\noindent
	© 2022 IEEE. Personal use of this material is permitted. Permission from IEEE must be obtained for all other uses, in any current or future media, including reprinting/republishing this material for advertising or promotional purposes, creating new collective works, for resale or redistribution to servers or lists, or reuse of any copyrighted component of this work in other works.
\end{titlepage}

\title{Anchor Pair Selection in TDOA Positioning Systems by Door Transition Error Minimization

\thanks{This research was partly supported by The Foundation for the Development of Radiocommunication and Multimedia Technologies and  the National Centre for Research and Development, Poland under grant PerMed/II/34/PerHeart/2022.}
}

\author{\IEEEauthorblockN{Marcin Kolakowski\IEEEauthorrefmark{1},
Jozef Modelski\IEEEauthorrefmark{2}}
\IEEEauthorblockA{Institute of Radioelectronics and Multimedia Technology\\
Warsaw University of Technology\\
Warsaw, Poland\\
Email: \IEEEauthorrefmark{1}marcin.kolakowski@pw.edu.pl,
\IEEEauthorrefmark{2}jozef.modelski@pw.edu.pl}
}

\maketitle

\begin{abstract}
This paper presents an adaptive anchor pairs selection algorithm for UWB (ultra-wideband) TDOA-based (Time Difference of Arrival) indoor positioning systems. The method assumes dividing the system operation area into zones. The most favorable anchor pairs are selected by minimizing the positioning errors in doorways leading to these zones where possible users' locations are limited to small, narrow areas. The sets are determined separately for going in and out of the zone to take users' body shadowing into account.
The determined anchor pairs are then used to calculate TDOA values and localize the user moving around the apartment with an Extended Kalman Filter based algorithm.

The method was tested experimentally in a furnished apartment. The results have shown that the adaptive selection of the anchor pairs leads to an increase in the user's localization accuracy. The median trajectory error was about 0.32 m.

\end{abstract}


\section{Introduction}
Indoor positioning is a rapidly developing field both in the case of conducted research and available commercial solutions. In the literature, there are multiple ultra-wideband (UWB) based solutions presented, which allow the potential users to obtain positioning results of high, decimeter-level accuracy.

Such high accuracy is usually obtained under specific Line-of-Sight (LOS) propagation conditions and for a vast number of anchor nodes. In most real scenarios, the infrastructure is limited, and the environments are cluttered with obstacles, making the system operate in Non-Line-of-Sight (NLoS) conditions most of the time. Maintaining high positioning accuracy requires implementing additional methods.

Most of the methods proposed in the literature consist in detecting measurement results obtained under NLOS conditions and mitigating their negative impact on accuracy by introducing appropriate corrections \cite{barralNLOSIdentificationMitigation2019}. The efficiency of such methods depends on accuracy of assumed NLOS bias statistics \cite{zouEfficientNLOSErrors2020}. Estimating them requires lengthy measurements which results, due to complex nature  of indoor environments, may not be transferable between various system deployment sites. A possible solution is to use one of the alternative anchor selection methods. Such methods are used to select measurement results, for which the obtained positioning accuracy would be best in terms of accuracy and reliability.

The following paper presents an anchor pairs selection method for Time Difference of Arrival (TDOA) positioning in UWB-based systems. The most favorable pairs for location in different zones of the system deployment area are established during a calibration routine by minimizing the positioning errors around doorways, where the user's location is known due to constraints imposed by narrow passages. The calibration is performed  using historical data comprising user locations and corresponding UWB Time of Arrival (TOA) measurements.

\section{Background and Related Works}
\begin{figure}[b]
\centerline{\includegraphics[width=.6\linewidth]{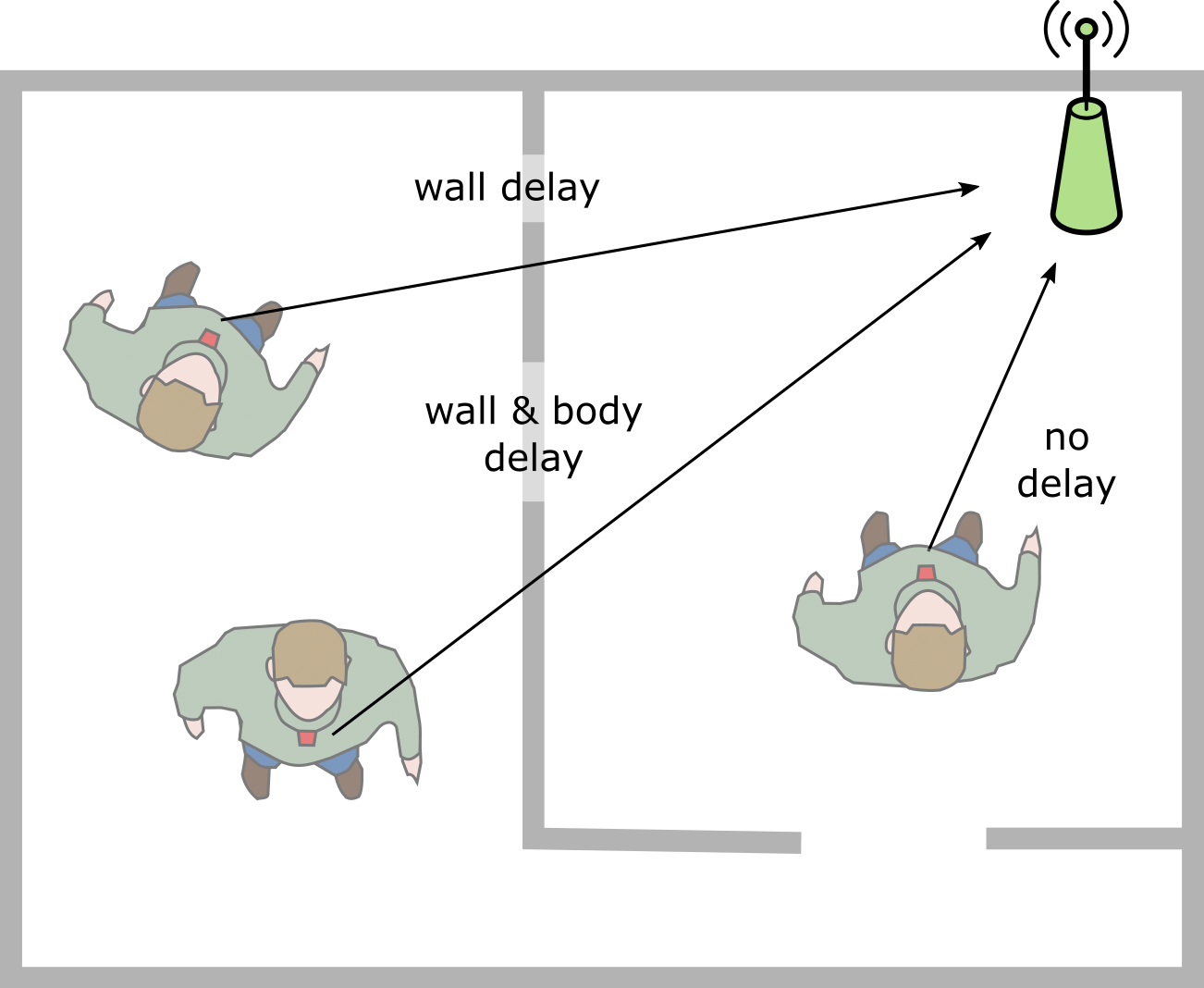}}
\caption{The NLOS delays introduced in indoor user localization scenario}
\label{fig:nlos_delays}
\end{figure}

Most of the time, localization systems deployed in indoor environments localize the users based on measurement results obtained in NLOS propagation conditions. The NLOS conditions can be caused by the stationary obstacles present where the system is installed (e.g., walls, pieces of furniture) and the user's body. The problem is illustrated in Fig. \ref{fig:nlos_delays}. 

The user moves through the environment and changes her/his orientation regarding the anchors comprising the system infrastructure. It creates NLOS conditions, which would be hard to mitigate due to multiple factors causing the delays. For example, the anchor might be located in a different room than the user, and through-wall propagation would delay the signal. When the user faces away from the anchor, there is an additional delay caused by body shadowing. 

The delay introduced by the body shadowing depends on the degree to which the user's body covers the tag. The study presented in \cite{otimFDTDEmpiricalExploration2019} has shown that the introduced UWB ranging delay rises sharply when the tag becomes wholly covered. In scenarios where the user's body only partially blocks the path between the tag and the anchor, the delays are usually moderate.

One way to reduce the impact of the introduced delays is to determine the most favorable measurement results set for each scenario (user's facing and the area, where he is located), and use only the results from selected anchors.

There are several approaches used for anchor selection in indoor positioning scenarios. The first is to choose only the anchors which are deemed to work in LOS conditions. It can be done using NLOS detection methods, which usually require prior training and calibration, or by establishing a criterion for anchor selection e.g., choosing the anchors based on the received signal strength or an estimated distance to the tag \cite{cantonpaternaBluetoothLowEnergy2017}. It assumes that for the anchors close to the tag, the probability of them being covered by an obstacle is low, which is not always accurate in a complex indoor environment.

The other approach consists in a preliminary analysis of the theoretical bounds of localization accuracy using Cramer-Rao Lower Bound (CRLB) \cite{daiNearlyOptimalSensor2020}. The studies have shown that the CRLB depends on the localized object's location and the employed set of anchor nodes\cite{kauneAccuracyStudiesTDOA}. Unfortunately, because CRLB analysis is applicable in scenarios where the measurement results are unbiased, it would be hard to use it in most indoor conditions.

The last group of methods depends on localization error minimization \cite{feilongOptimumReferenceNode2015,monicaUWBbasedLocalizationLarge2015, kolakowskiAdaptiveAnchorPairs2021}.  They consist in localizing the tag based on multiple measurement sets and choosing the best one by comparing the results with the ground truth. This approach yields good results, but its use requires prior system calibration either by manual measurements or using a robotic platform.

The method proposed in the paper adopts the above approach. The presented research is a continuation of works described in \cite{kolakowskiAdaptiveAnchorPairs2021}, where the best anchor sets were determined through minimization of a moving robot positioning errors. The conducted experiments have shown that using the proposed solution results in a significant improvement in localization accuracy. However, that method has two significant drawbacks. First of all, the calibration routine requires using a robotic platform with self-location abilities. Otherwise, the data would need to be collected manually, which is a time-consuming process. Secondly, the calibration is performed for a robot rather than a system user, and thus, the results do not take body shadowing into account. 

The method proposed in the paper solves the above problems. The calibration is performed based on the data gathered during routine system operation and therefore accounts for additional disturbances caused by the user's body.

\section{Proposed Method}
\subsection{Method concept}

\begin{figure}[t]
\centerline{\includegraphics[width=.8\linewidth]{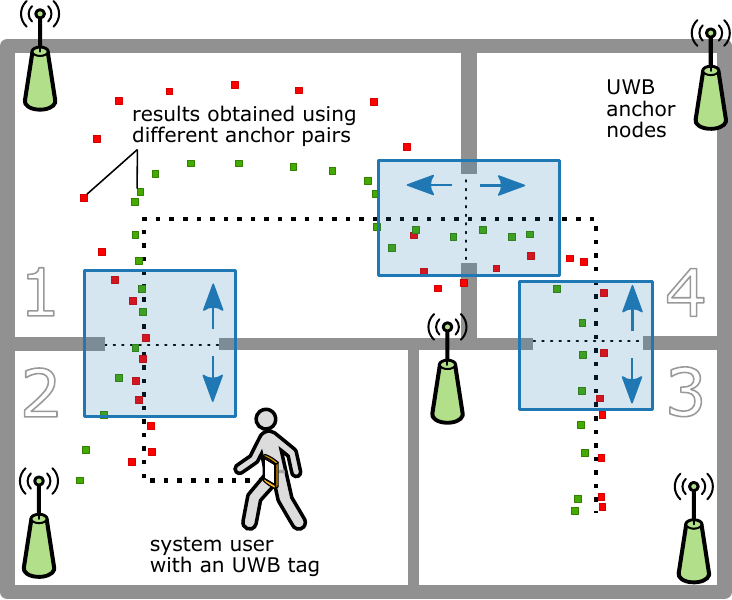}}
\caption{The proposed method's concept}
\label{fig:concept}
\end{figure}

The concept of the proposed anchor selection method is presented in Fig. \ref{fig:concept}.
The proposed method divides the environment into zones and determines the most favorable anchor pairs sets for each of them. In a typical scenario, the user uses the system as usual, and after some time, e.g. at the end of the day, the system uses the gathered data to recalculate users locations using different sets of TDOA values and choose the ones, which allow localizing the user with the highest accuracy.

\begin{figure}[t]
\centerline{\includegraphics[width=.7\linewidth]{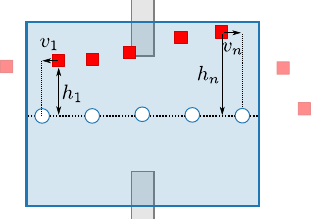}}
\caption{Cost metrics}
\label{fig:cost}
\end{figure}

The positioning accuracy for each anchor set is done by comparing the obtained results against environmental constraints imposed on the possible user's locations by narrow doorways leading to the zones. In the proposed method, the results accuracy is rated based on the cost function: 
\begin{equation}
cost(\mathrm{TDOAs})=\sum_{i=1}^n h_i + v_i
\label{eq:metric}
\end{equation}
Where $n$ is the total number of points used for accuracy evaluation and $h_i$ and $v_i$ are distances presented in Fig. \ref{fig:cost}. The  $h_i$ and $v_i$ distances are projections of vectors between positioning results and reference points, which are equally distributed on the door axis and represent the user's location when moving through the doorway.

The accuracy analysis is performed for both door transition directions. Two anchor pair sets corresponding to opposite user headings are determined, which helps to reduce body impact on positioning accuracy. In some situations, there will be several anchor sets for a single zone (e.g., zone 1 will have four anchor sets assigned - one for each user heading).

\subsection{Door detection}
\begin{figure}[b]
\centerline{\includegraphics[width=\linewidth]{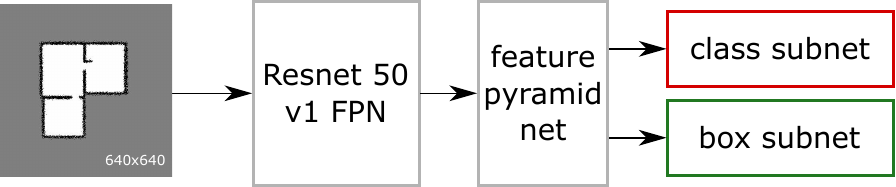}}
\caption{Door detector architecture}
\label{fig:NN_arch}
\end{figure}

The proposed method requires knowledge of door locations. In the case of small spaces, where the number of doors is limited, this information can be manually extracted from construction plans. However, it would be best to automate the process. In the proposed method, door detection is performed automatically using a Single Shot Detector based on RetinaNet\cite{RetinanetSSDResnet} architecture with ResNet50 v1 FPN feature extractor. The architecture of the proposed solution is presented in Fig. \ref{fig:NN_arch}.

The network takes 640x640 pixel images containing occupancy grid maps of the environments as an input. The resolution of the grid map should be 5 cm. The image passes through feature extraction layers, and then doors are detected, and their bounding boxes are estimated.

\begin{figure}[t]
\centerline{\includegraphics[width=.9\linewidth]{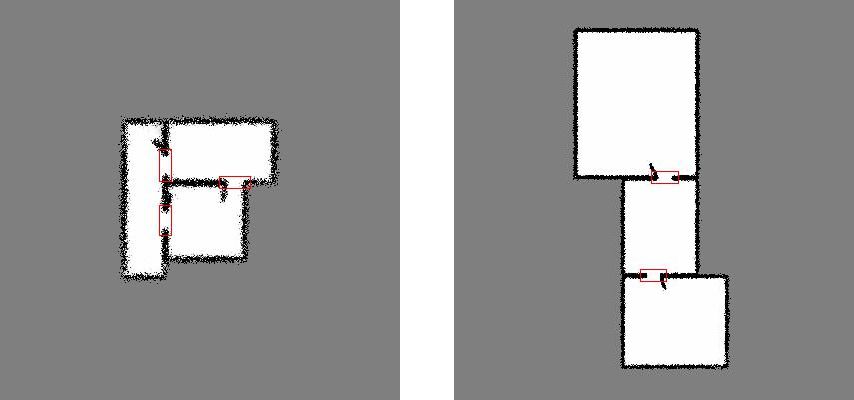}}
\caption{Training examples with doorway boxes marked}
\label{fig:test_examples}
\end{figure}

In our study, we have used transfer learning. The feature extractor weights were obtained from the Tensorflow model garden, and we fine-tuned only the classification and regression networks using a custom training set containing random grid maps. The grid maps were generated using Python scripts in such a manner that each image contained at least two doorways. The doorways were of different widths (from 0.6 to 0.8 m), and the grid maps were noisy to make them more similar to the LiDAR obtained ones. Exemplary training images with doors marked are presented in Fig. \ref{fig:test_examples}.

\subsection{Positioning algorithm}

\begin{figure}[t]
\centerline{\includegraphics[width=.7\linewidth]{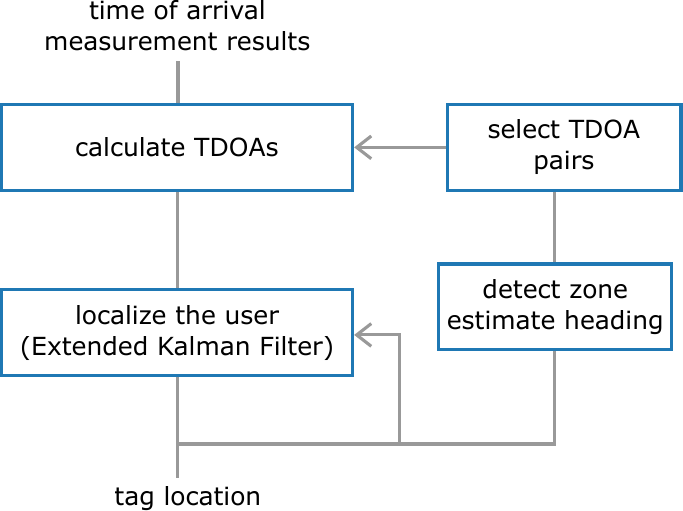}}
\caption{The proposed localization algorithm workflow}
\label{fig:algorithm}
\end{figure}

The workflow of the proposed localization algorithm is presented in Fig. \ref{fig:algorithm}. The algorithm consists of two steps. First, the times of arrival values measured by the infrastructure are used to calculate the TDOA values. The TDOA pairs to be calculated are chosen in the feedback loop, in which the current user location (with respect to the zone) and heading are estimated. The obtained TDOAs are used to localize the user using an Extended Kalman Filter-based algorithm.

\section{Experiments}
\begin{figure}[!tbhp]
\centerline{\includegraphics[width=.92\linewidth]{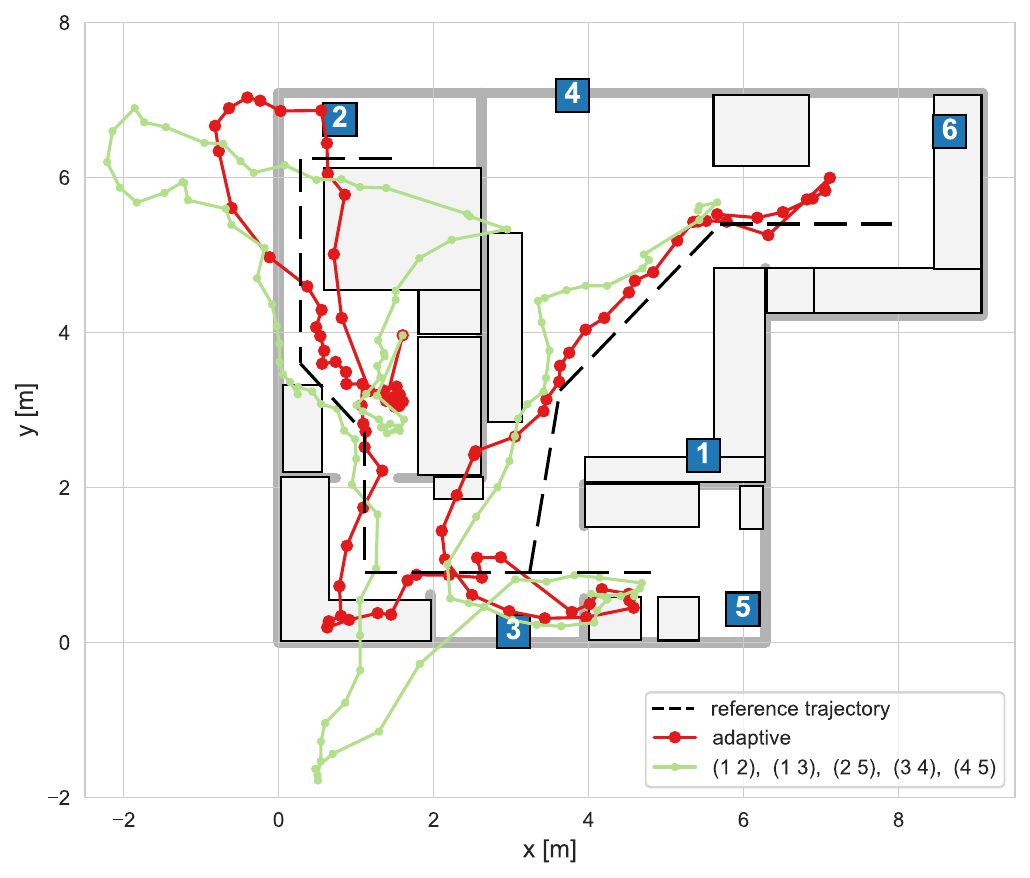}}
\caption{Positioning results. The blue squares represent anchor positions. Gray rectangles represent obstacles, which block the users movement (bed, couch etc.).}
\label{fig:exp_res}
\end{figure}

The experiments were conducted in a fully furnished, 48 sq m apartment, which layout is presented in Fig. \ref{fig:exp_res}. The system used in the study \cite{kolakowskiUWBBLETracking2020} consisted of 6 anchors and a tag worn by the user on a lanyard. The test consisted of walking along a test path and visiting all apartment rooms. 

\begin{figure}[t]
\centerline{\includegraphics[width=.6\linewidth]{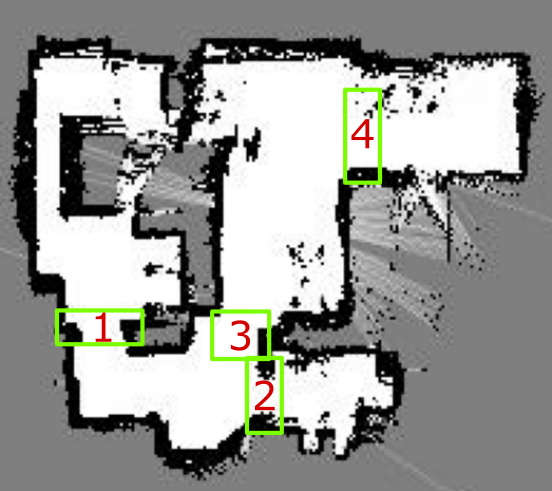}}
\caption{Door detection results marked on an LiDAR based occupancy grid map}
\label{fig:door_detection}
\end{figure}

The data processing started with detecting the doors on the occupancy grid map obtained with a LiDAR. The detector was implemented using TensorFlow Object Detection API. The results of door detection are presented in Fig. \ref{fig:door_detection}.

In the test environment, four doorways were detected. The network detected doorways with proper door frames and other doorway-like narrow spaces, e.g., a passage to the kitchen between a wall and a table (marked with 4 in Fig. \ref{fig:door_detection}).

Then, an extensive set of measurement data corresponding to three hours of constant walking in the apartment was simulated. The simulated measurements took into account the delay resulting from through the wall propagation (0.3 ns per traversed wall) and the one due to the user's body shadowing modeled based on the results presented in \cite{otimFDTDEmpiricalExploration2019}. The simulated data were used to establish the most favorable anchor sets for each of the zones (listed in Table \ref{tab:best_pairs}).

\begin{table}[t]
\caption{Most favorable sets of TDOA pairs (based on simulated data)}
\begin{center}
\def\arraystretch{1.5}
\begin{tabular}{c|c}

\textbf{zone}&\textbf{anchor pairs}\\
\hline
1$\uparrow$&(1 2),  (1 6),  (3 5),  (5 6)\\
1$\downarrow$&(1 3),  (1 6),  (2 5),  (3 4),  (3 6),  (4 5)\\
2$\rightarrow$&(1 3),  (1 6),  (2 4),  (3 6),  (4 6)\\
2$\leftarrow$&(1 3),  (1 5),  (2 4),  (3 6)\\
3$\uparrow$&(1 4),  (1 6),  (2 3),  (2 6),  (3 5),  (4 6)\\
3$\downarrow$&(1 6),  (2 3),  (2 5),  (3 5),  (3 6),  (4 6)\\
4$\rightarrow$&(1 2),  (1 3),  (1 4),  (2 4),  (4 6)\\
4$\leftarrow$&(1 4),  (1 5),  (2 4),  (2 6),  (3 4),  (3 5)\\
\end{tabular}
\label{tab:best_pairs}
\end{center}
\end{table}

\begin{figure}[htbp]
\centerline{\includegraphics[width=.8\linewidth]{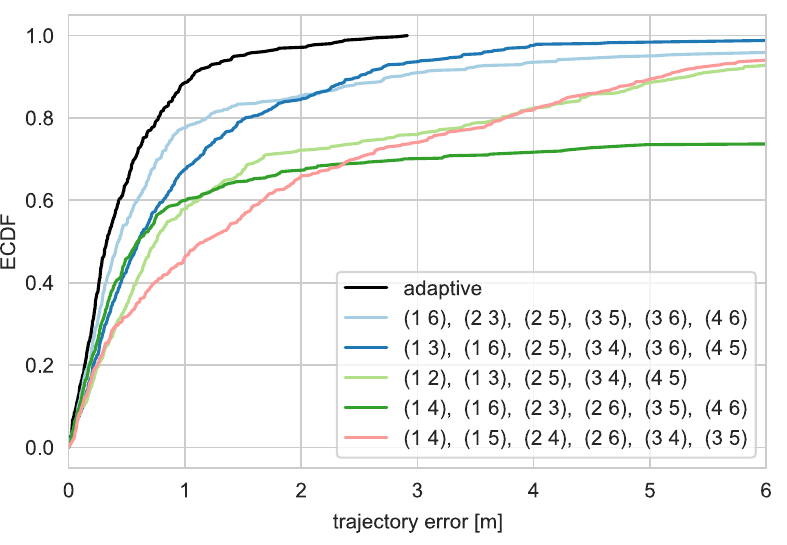}}
\caption{Estimated Cumulative Distribution Function of trajectory error for different sets of TDOA pairs}
\label{fig:cdfs}
\end{figure}

The obtained pairs were then used to localize the user walking along the test trajectory. Exemplary positioning results obtained using the proposed method and one fixed TDOA pair set are shown in Fig.\ref{fig:exp_res}. The Estimated Cumulative Distribution function of trajectory error, defined as the smallest distance of positioning results from reference trajectory are presented in Fig. \ref{fig:cdfs}.

The obtained results allow reconstruction of the user's trajectory with satisfying accuracy. Using different sets for different areas makes it possible to localize the user in all rooms accurately. In the case of using a fixed TDOA pair set, it was impossible to maintain high accuracy in all of the spaces (e.g., localization accuracy in the room on the left is much lower). The mean trajectory error of the proposed method was about 0.32 m which is, on average, about 0.4 m better than in the case of using one set for the whole apartment, which is a similar gain to the one reported based on simulations in \cite{feilongOptimumReferenceNode2015}.

\section{Conclusions}
The paper presents a novel method of TDOA pairs selection intended for UWB systems. The method determines the most favorable anchor sets by minimizing the localization error obtained with different results combinations. Instead of ground truth data, the method uses environmental constraints imposed on possible users' locations by narrow passages (e.g., doors).

The performed experiments have shown that the method allows the system to maintain high accuracy in all covered areas, which was not the case while using fixed anchors sets.

The method can be further developed by improving the door detection network and including other narrow passages (e.g., created by furniture pieces and the walls) in the analysis.


\vspace{12pt}

\end{document}